\documentclass{JHEP3}
\let\ifpdf\relax

\pdfoutput=1

\usepackage[parfill]{parskip}
\usepackage{graphicx}
\usepackage{fancyvrb}
\usepackage{makeidx}
\usepackage{latexsym,amssymb,amsmath,graphicx, makeidx}
\usepackage{ifpdf}

\newcommand{\be}{\begin{equation}}
\DefineVerbatimEnvironment{code}{Verbatim}{fontsize=\small}
\newcommand{\ee}{\end{equation}}
\newcommand{\bi}{\begin{itemize}}
\newcommand{\ei}{\end{itemize}}
\makeindex

\newcommand{\ldd}{\lambda_{\Delta}}
\newcommand{\ldc}{\lambda_{\Delta}^c}
\newcommand{\lhh}{\lambda_{H}}
\newcommand{\lddc}{\lambda_{\Delta\Delta^c}}

\preprint{PI-PARTPHYS-281}

\title{Dark Resonance}

\author{ Haipeng An$^a$, Maxim Pospelov$^{a,b}$ \\
        $^a$Perimeter Institute for Theoretical Physics, Waterloo, ON, N2J 2W9, Canada\\
        $^b$Department of Physics and Astronomy, University of Victoria, Victoria, BC, V8P 5C2 Canada  \\
       E-mail: \email{han@perimeterinstitute.ca, mpospelov@perimeterinstitute.ca}
               }

\abstract{

We construct explicit models of particle dark matter where the attractive force in the dark matter sector 
creates a narrow near-threshold resonance that qualitatively changes the energy dependence of the
annihilation cross section. In these models, the resonant 
enhancement of the dark matter annihilation can easily 
source the excess of energetic leptons observed by experiments on PAMELA and FERMI satellites. 
The distinct feature of these models is that by construction 
the enhancement of the annihilation cross section shuts off when the dark matter 
velocity falls below the typical Milky Way values, thus  automatically satisfying constraints  
on dark matter annihilation imposed by the CMB anisotropies and gamma ray constraints from 
satellite galaxies. 
}

\begin{document}
\maketitle
\tableofcontents

\section{Introduction}

It is well-known that baryonic matter contributes only about 20\% of the mass density in the Universe. A plethora of independent cosmological observations strongly support that the remaining 80\% is composed of cold dark matter (DM). The most recent observation suggests that DM contributes a fraction of $\Omega_{\rm DM} = 0.229\pm0.015$ to the total energy content of the universe~\cite{Komatsu:2010fb}. 
Notwithstanding our rather precise knowledge of the global dark matter 
energy density, its nature remains a mystery, and constitutes one of the most profound questions in fundamental 
physics. Several candidates have been proposed to account for DM~\cite{Feng:2010gw}, 
with weakly-interacting massive particle (WIMP) featured prominently among them, 
thanks to the natural mechanism for the its relic abundance regulated 
through thermal annihilation cross section of about $\langle\sigma v\rangle \approx 3\times10^{-26} ~{\rm cm^{2} s^{-1}}$. The annihilation of DM into the Standard Model (SM) particles in the local Universe, and specifically inside our galaxy, can result in  its``indirect detection'' via energetic annihilation products such as 
high energy gamma rays, charged leptons, protons and and anti-protons. 

Recent measurements of cosmic ray spectra of charged leptons 
by the PAMELA, ATIC, Fermi and H.E.S.S experiments have reported 
some "anomalies" relative to the prior expectations based on theoretical models of 
cosmic rays. 
In particular, the PAMELA results show a rise in the relative contribution of positron flux into the 
total $e^+e^-$ flux above energies of around 10 GeV to 100 GeV~\cite{Adriani:2008zr}.  Also, a broad excess in the total $e^+e^-$ spectrum at around TeV scale energies was reported by several experiments~\cite{Chang:2008aa,Abdo:2009zk,Aharonian:2009ah}. 
On the other hand, the flux of antiprotons does not show any anomaly relative to prior expectations \cite{Adriani:2010rc}.
While it is possible that some unaccounted astrophysical sources are behind the excess of energetic leptons, 
the DM explanation to these excesses has also being entertained, steering theoretical modifications of the 
minimal WIMP paradigm. 
These excesses are not easily accommodated in the thermal WIMP scenario, because of the need to enhance the annihilation 
cross section into leptons in the galactic environment by a large factor $O(10^2-10^3)$, while maintaing  the antiproton fraction within limits. 
Both model-building goals, enhancement of the cross section, and suppression of antiprotons is naturally achieved 
in models that supplies the DM with the the so called "dark force", a relatively light compared to the WIMP mass mediator 
particle that connects DM with the SM~\cite{ArkaniHamed:2008qn,Pospelov:2008jd,Baumgart:2009tn,Katz:2009qq}.
The enhancement of the annihilation rate in galactic environment can be broadly described as 
due to the extra mediator-exchange attractive force  that creates the Coulomb (or Sommerfeld) enhancement of the cross sections at low WIMP velocities. Ref. \cite{Pospelov:2008jd} summarizes certain broad classes of enhancement mechanisms within this scenario: 
1. the usual Sommerfeld type enhancement that can be approximated as ${\cal S}_{\rm \rm Coul} \simeq  \pi \alpha'/v$ in the 
limit of small mass of the mediator, with $v$ being WIMP relative velocity and $\alpha'$ the coupling constant of the dark force; 2. narrow resonances  that can bring the enhancement far above ${\cal S}_{\rm \rm Coul}$; 3. recombination into WIMP-anti-WIMP bound state with the emission of a mediator, that brings further numerical enhancement to the 
cross section. 

Further developments included an in-depth calculations of the Sommerfeld enhancement as the function of the mass of the
mediator and possible small mass splittings of WIMP states  \cite{Slatyer:2009vg}, and a
discussion on whether these models afford the required enhancement factors \cite{Feng:2009hw,Finkbeiner:2010sm,Cirelli:2010nh,Pospelov:2010cw}.
It is fair to say at this point that  simple versions of such scenarios are under strong pressure
from precision data on anisotropies of the cosmic microwave background (CMB) from WMAP7~\cite{Galli:2011rz,Finkbeiner:2011dx}, and the (non)observation of gamma rays from dwarf spheroidal galaxies by Fermi-LAT~\cite{Ackermann:2011wa,GeringerSameth:2011iw,Mazziotta:2012ux,Cholis:2012am,Essig:2010em}.
Neither of these latter studies have so far  given any indication on an enhanced DM annihilation signal. 
However, the kinetic energy of  WIMP particles in dwarf spheroidals and in the universe during the 
CMB decoupling is much smaller than the average kinetic energy of DM in our galaxy. 
The momentum of DM redshifts in the same rate as the expansion of the Universe after its thermal decoupling. 
Therefore,  the average kinetic energy of DM at the recombination 
must be smaller than the temperature of the universe, which is about 0.3 eV. Inside dwarf spheroidal galaxies, the velocity dispersion is around or smaller than 10 km s$^{-1}$~\cite{Walker:2009zp,Koposov:2011zi}, which is one order of magnitude smaller than 200 km s$^{-1}$,
typical for DM velocities in our galaxy in the $\sim$ kpc vicinity of the Solar System. 
Therefore, {\em if} the enhancement factor depends on the kinetic energy of dark matter in a non-monotonic way and peaks when the relative velocity of DM is around 200 km s$^{-1}$,
 it can co-exist with  constraints from the CMB and  dwarf spheroidal galaxies, and yet generate the desired enhancement aimed at explaining the anomalous spectra of positrons and electrons with DM annihilation. 

The purpose of our work is to show 
that the existence of narrow near-threshold resonances in WIMP-WIMP system can easily invalidate the extrapolation of galactic annihilation rate to lower velocities, enabling to escape constraints from the CMB and dwarf spheroidals. The main idea of our work is illustrated pictorially in Fig. 1. The narrow 
resonance (black curve pointing upward at 1 MeV, $\Gamma \sim 0.1$ keV) 
maximizes the overlap with DM kinetic 
energy distribution inside the galaxy, represented on the plot by the Maxwell distribution with $T=1$ MeV. 
At the same time, the overlap with the early universe distribution (blue curve), and less energetic 
distribution for dwarf spheroidal galaxies (green cure, $T=0.01$ MeV) is minimized. This should be 
contrasted with the case of a typical Sommerfeld-enhanced cross section (black curve with flat behaviour
at $E\to 0$,  Fig. 1) that has significant overlap with both red and green distributions. 

\FIGURE[h!]{
\includegraphics[scale=1]{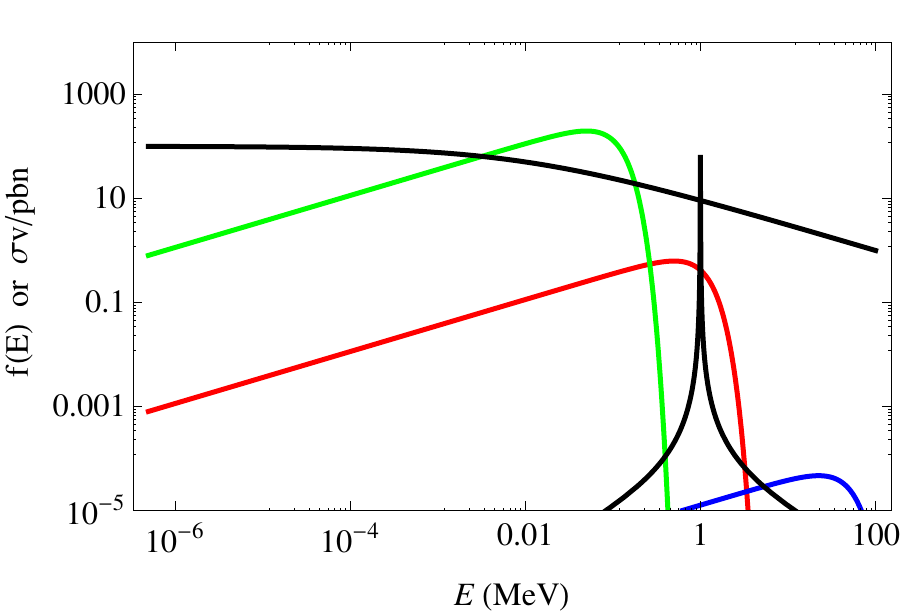}
\caption{Schematic illustration of the main idea about the importance of 
narrow near-threshold resonance for the DM phenomenology. In black, are the  $\sigma v$ in units of pb 
for a generic Sommerfeld-enhanced annihilation with the saturation at low energies and the narrow
$p$-wave resonant cross section with the peak at $E=1$ MeV. The annihilation rate is given by the integral from the overlap of $\sigma v$ with the DM energy distribution, which are pictured as Maxwell distributions 
for energetic particles in the early universe (blue), galactic environment (red) and sub-haloes (green). 
Narrow resonance can have a large overlap with galactic DM distribution, while minimizing the rate both for 
large and small velocity dispersions. Sommerfeld enhanced model is guaranteed to have the enhancement 
for the small velocity dispersions.
\label{fig:idea}}}

A general formula for the resonance-enhanced scattering cross section can be written as 
\begin{equation}\label{master}
\sigma = \frac{2J+1}{(2S_1 + 1)(2S_2+1)}~\frac{\pi}{k^2}~ \frac{\Gamma_{\rm in}(E) \Gamma_{\rm out}}{(E-E_R)^2 + \Gamma_{\rm tot}^2(E)/4}\ ,
\end{equation}
where $k$ is the momentum in the centre-of-mass frame, $J$ is the total angular momentum of the resonance, $S_1$ and $S_2$ are the spins of two incoming particles, $E$ is the centre-of-mass energy, $\Gamma_{\rm in}$ and $\Gamma_{\rm out}$ are incoming and outgoing widths. Notice that 
since the resonance is close to the zero-energy threshold, the 
entrance width must depend on the momentum of the particles and can be parameterized as 
$\Gamma_{\rm in}(E) = \Gamma_{\rm in}^{(0)} (E/E_R)^{l+1/2}$, where  $
l$ is the angular momentum of the system. In the limit of the very narrow resonance the energy 
dependence of Eq. (\ref{master}) is often approximated by the delta-function.

The origin of the resonances may vary. It can be generated by the intermediate $s$-channel 
mediator with the mass closely tuned to $2m_{\rm WIMP}$, Ref.~\cite{Ibe:2008ye}. A more natural, 
{\em i.e.} less tuned, scenario might be that two DM particles form a resonance through a long range attractive force \cite{Pospelov:2008qx,Pospelov:2008jd,Backovic:2009rw}. While the standard Sommerfeld enhancement 
models do exhibit a series of resonances (see {\em e.g.} \cite{Slatyer:2009vg}), these are not suitable for our purposes 
because of their large width. Therefore, a dedicated  exercise is required in order to 
find whether scenario in Fig. 1 can have concrete model-building realizations. 
In this paper, we propose two prototype models, which can induce an $O(10^3)$ resonant 
enhancement factor for the DM annihilation in our local galaxy, while  avoiding 
constraints from the CMB and dwarf spheroidal galaxies. We view these models as a "proof of existence":
not overly complicated models of dark sector can indeed furnish the required non-monotonic behaviour of the 
enhancement factor, which calls for more in-depth analysis of phenomenology of such models. 

The rest of the paper will be organized as  follows: In Sec. 2 and Sec. 3 we propose two candidate models that have narrow resonances either in $s$ or $p$ wave. In Sec. 4 we provide a discussion of the 
gamma-ray constraints from the galactic centre. We summarize the results in Sec. 5.   

\section{Narrow resonance in the $s$-wave}

\subsection{The model}

The main idea of the model is to realize the $s$-wave resonance via the small mass splitting of two WIMP mass eigenstates, 
$\chi_1$ and $\chi_2$, with $\chi_2$ being the lightest and stable. 
The following annihilation chain is envisaged:
\begin{equation}
\label{process1}
\chi_2+\chi_2 \to (\chi_1 \chi_2)_{1s~\rm resonance} \to {\rm light ~ mediators} \to {\rm SM}.
\end{equation}
While our models can be constructed for different spin of dark matter particles, we prefer to work with scalars as the most economical and
simplest option. 

We choose the Lagrangian of the dark sector to consist of new $U(1)$ gauge group, one complex DM scalar $\chi$ and 
three sets of Higgs fields, $H$, $\Delta$ and $\Delta^c$:
\begin{equation}
{\cal L} = {\cal L}_0 - \frac{1}{4}(\lhh {\chi^\dagger}^2 H^2 + {\rm h.c.}) - \frac{1}{2}(\ldd \Delta^\dagger \chi^2 + {\rm h.c.}) - \frac{1}{2}(\ldc{\Delta^c}^\dagger \chi^2 + {\rm h.c.}) - \frac{1}{4}(\lddc \Delta^2 {\Delta^c}^2 + {\rm h.c.}) \ ,
\end{equation}
where ${\cal L}_0$ includes  the SM Lagrangian, the new $U(1) $ field Lagrangian as well as other terms that  are not crucial for the DM phenomenology. 
Under the dark $U(1)$ symmetry  $\chi$ and $H$ are singly charged, while  $\Delta$ and $\Delta^c$ cary the charge 
$\pm2$. We assume that this dark $U(1)$ symmetry is spontaneously broken by the vacuum expectation value (vev) of 
$\Delta^c$. Without loss of generality, we assume coupling constants $\lddc$, $\ldc$ and $\lhh$ to be real and leave only $\ldd$ to be complex. Then after spontaneous symmetry breaking, the real and imaginary parts of $\chi$ and $\Delta$ are split into the mass eigenstates, 
with eigenvalues given by 
\begin{eqnarray}
m^2_{\chi_1} = m_\chi^2 + \frac{1}{2\sqrt{2}}\ldc v \ , &&\;\; m^2_{\chi_2} = m_\chi^2 - \frac{1}{2\sqrt{2}}\ldc v \nonumber\\
m^2_{\Delta_1} = m_\Delta^2 + \frac{1}{4}\lddc v \ , &&\;\; m^2_{\Delta_2} = m_\chi^2 - \frac{1}{4}\lddc v \nonumber\\
\end{eqnarray}

The residual interaction Lagrangian between $\chi$ and $\Delta$ can be written as
\begin{eqnarray}
\label{Higgsint}
{\cal L}_{\chi\Delta} &=& -\frac{1}{\sqrt{2}}\Delta_1 \left[ \frac{{\rm Re}\ldd}{2}(\chi_1^2-\chi_2^2) - {\rm Im}\ldd \chi_1 \chi_2 \right] \nonumber\\
&&- \frac{1}{\sqrt{2}}\Delta_2 \left[ \frac{{\rm Im}\ldd}{2}(\chi_1^2-\chi_2^2) + {\rm Re}\ldd \chi_1 \chi_2 \right]\ .
\end{eqnarray}

In the primordial Universe, the  thermal annihilation of dark matter stops when the temperature 
drops to a level $T_{\rm stop} \sim M_\chi/20$. The mass difference between the real and imaginary parts of $\chi$ is negligible compared to $T_{\rm stop}$, and one can treat $\chi_1$ and $\chi_2$ as one complex scalar. To discuss the thermal annihilation, let us consider the case with ${\rm Re}\lambda_H 
\sim {\rm Im}\lambda_H \gg $ other four-point couplings. 
Then, there are two main annihilation channels,  $H$ scalars and vector bosons, with the following cross sections:
\begin{eqnarray}
&&(\sigma v)_{2\chi \rightarrow 2H} = (\sigma v)_{{2\chi^\dagger} \rightarrow 2H^\dagger } \approx 
\frac{|\lambda_H|^2}{32\pi M_\chi^2}  \ ,\nonumber \\
&&(\sigma v)_{\chi\chi^\dagger \rightarrow V^2} \approx \frac{\pi \alpha_V^2}{M_\chi^2}\ ,
\end{eqnarray}
and its sum will determine the DM abundance. For the choice of parameters considered below, the $H$ channel is the dominant 
one.

\subsection{Formation of the bound state}

The ladder exchange by $V_\mu$ leads to the attractive force between $\chi_1$ and $\chi_2$ particles, Fig. 
\ref{fig:ladder}. It also enables the attraction in the $\chi_1\chi_1 \leftrightarrow \chi_2\chi_2$ channel. 
One can readily observe that $V$-exchange respects certain charge symmetry in $\chi_i\chi_j$ sector:
$i+j$-even and $i+j$-odd sectors do not mix until one includes Higgs-induced interactions (\ref{Higgsint}). 
The attractive $V$-mediated force leads to the bound states in both sectors, but 
by increasing the mass of $V$ or reducing the couplings,
 the bound states can be squeezed above $\chi_2\chi_2$ separation 
threshold. 
We choose the parameters in such a way that $\chi_1\chi_2$ state is just above the threshold.

Let us consider the first chain in process (\ref{process1}),  namely the capture into $(\chi_1\chi_2)_{1s}$ 
resonance. From ${\cal L}_{\chi\Delta}$ one can see that 
both $\Delta_1$ and $\Delta_2$ induce a process of $\chi_2 \chi_2 \rightarrow \chi_1\chi_2$. 
Since in the absence of the Higgs-mediated interaction the two sectors, $\chi_2\chi_2$ and $\chi_1\chi_2$ 
do not mix, it is rather easy to obtain a very small entrance width for this resonance. 
After some lengthy 
but straightforward calculations,
we derive
\begin{equation}\label{gamma_in}
\Gamma_{\rm in}(v_r) = \frac{{\rm Re}\ldd^2{\rm Im}\ldd^2 a_B v_r}{64 \pi^2 M_\chi^2 } \left[ \frac{1}{(1 + m_{\Delta_1} a_B)^2} - \frac{1}{(1 + m_{\Delta_2} a_B)^2} \right]^2\ ,
\end{equation}
where $a_B = 2\pi/(\alpha_V M_{\chi})$ is the Bohr radius of the bound state and $v_r$ is the relative velocity between the two $\chi_2$ 
particles in the continuum. 

We next calculated $\Gamma_{\rm out}$ in our model. It turns out that $\chi_1\chi_2$ pair annihilation into one gauge boson 
and one scalar in the final state is negligible.  (It is either in $p$-wave of is additionally 
suppressed by a small vev). Since the annihilation happens at short distances,
one can conveniently tie $\Gamma_{\rm out}$ to the annihilation cross section of $\chi_1\chi_2\rightarrow H$ scalars:
\begin{equation}
(\sigma v)_{\chi_1\chi_2\to HH} \equiv \langle \sigma v\rangle_0 \approx \frac{\lhh^2}{32\pi M_\chi^2}  \ ,
\end{equation}	
so that the out-going width becomes 
\begin{equation}
\Gamma_{\rm out} \approx \langle \sigma v\rangle_0 4\pi|\psi(0)|^2 = \frac{\lhh^2}{8\pi M_\chi^2 a_B^3}  \ . 
\end{equation}

It is important to specify the ratio of two widths in our model:
\begin{equation}
\frac{\Gamma_{\rm in}}{\Gamma_{\rm out}} = \frac{2 v_{r}}{\pi} \frac{{\rm Re}\ldd^2{\rm Im}\ldd^2}{\lhh^2 (\alpha_V M_\chi)^4} \left[ \frac{1}{(1 + m_{\Delta_1} a_B)^2} - \frac{1}{(1 + m_{\Delta_2} a_B)^2} \right]^2 \ .
\end{equation}
In order to have a large enhancement and a narrow width at the same time one would like to have 
\begin{equation}
\Gamma_{\rm in} \sim \Gamma_{\rm out} \sim {\rm ~keV}\ ,
\end{equation}
which removes a lot of freedom in the choice of parameters.
Substituting explicit expressions for $a_B$ into $\Gamma_{\rm out}$, we have
\begin{equation}
\Gamma_{\rm out} \approx \frac{1}{2} \alpha_V^3 M_\chi^3 \langle \sigma v\rangle_0 \approx \alpha_V^3 ~{\rm GeV}\ .
\end{equation}
For the WIMP mass of 1 TeV, the desired value for $\Gamma_{\rm out}$ is achieved with 
$\alpha_V \sim 0.01$, so that $\alpha_V M_\chi \sim 10$ GeV and $a_B \sim (5~ {\rm GeV})^{-1}$. 
Moreover, the requirement of the primordial annihilation rate to be consistent with the 
present day DM energy density, $\langle \sigma v\rangle_0\approx 1 ~{\rm pb}$, further 
imposes the requirement of $\lhh \sim 1$. 
Therefore, to make $\Gamma_{\rm in} \approx \Gamma_{\rm out}$ we need 
\begin{equation}
\left[\frac{({{\rm Re}\ldd~{\rm Im}\ldd})^{1/2}}{10~{\rm GeV}}\right]^4 \left[ \frac{1}{(1 + m_{\Delta_1} a_B)^2} - \frac{1}{(1 + m_{\Delta_2} a_B)^2} \right]^2 \sim 10^3\ .
\end{equation}

Since $\Delta_1$ and $\Delta_2$ both mediate an attractive interaction for the  $\chi_2$ pair, we have to make sure that 
this interaction is not generating strong Sommerfeld-type enhancement 
so that the model cannot be ruled out by 
satellite galaxies and CMB constraints. In the region $m_{\Delta_{(1,2)}} \gg 5 ~{\rm GeV}$ and $m_{\Delta_1} \gg m_{\Delta_2}$ 
the above relation can be rewritten as 
\begin{equation}
\left(\frac{\ldd}{2m_{\Delta_2} }\right)^4 \sim 10^3 \ . 
\end{equation}
If one  defines a dimensionless coupling in the following way 
\begin{equation}
\alpha_\Delta = \frac{1}{4\pi}\left(\frac{\ldd}{2\sqrt{2} M_\chi}\right)^2 \ , 
\end{equation}
the Sommerfeld enhancement factor generated by $\Delta_2$ can be written as
\begin{equation}
S_\Delta = \frac{\pi \alpha_\Delta M_\chi}{m_{\Delta_2}} = \frac{\ldd}{2m_{\Delta_2}}\frac{\ldd}{16 M_\chi} \approx \frac{0.3 \ldd}{M_\chi} \ .
\end{equation}
Therefore, one can readily see that as long as $\ldd < M_{\chi}$, $\Delta_{2}$-exchange 
 will not generate any enhancement in the annihilation of two $\chi_{2}$ particles. 

We now address the Sommerfeld enhancement due to exchange of $V$, as 
shown in Fig.~\ref{fig:ladder}(b). The effective potential for $|\chi_1\chi_1\rangle$ and $|\chi_2\chi_2\rangle$ system can be written as~\cite{ArkaniHamed:2008qn,Slatyer:2009vg}
\begin{equation}
{\cal V} = \left( \begin{array}{cc} 0 ~&~ \frac{\alpha e^{-m_V r}}{r} \\ \frac{\alpha e^{-m_V r}}{r} ~&~ 2 (M_{\chi_2} - M_{\chi_1})\end{array} \right) \ .
\end{equation}
The potential is attractive for one of the combination of $|\chi_1\chi_1\rangle$ and $|\chi_2\chi_2\rangle$ and repulsive for the other. However, in the case that $2(M_{\chi_1}-M_{\chi_2}) > E_{\rm binding}$, where $E_{\rm binding}$ is the binding energy of the 1$s$ bound state, one cannot get a sizeable Sommerfeld enhancement. 

Picking up a 
specific point in the parameter space with $M_{\chi_2} - M_{\chi_1} = 8$ MeV ($\epsilon_\delta\equiv\sqrt{2(M_{\chi_2}-M_{\chi_1})/M_{\chi}} / \alpha_V \approx 0.4$), $E_{\rm binding} = 7$ MeV, and $m_V = 2.6$ GeV, in Fig.~\ref{fig:sigma_v1} we show the dark matter annihilation cross section as a function of the kinetic energy $E_k$, including both 
 $\chi_2\chi_2$ and $\chi_1\chi_2$ intermediate channels. 
The blue dotted curve is the contribution from the $\chi_1\chi_2$ resonance, while the green dashed curve is the contribution from the Sommerfeld enhanced $\chi_2\chi_2$ annihilation.    

Assuming the distribution over kinetic energy to be Maxwellian, we also plot 
$\langle\sigma v\rangle$ as a function of temperature, Fig.~\ref{fig:sigma_va1}. 
One can readily observe the desired behaviour anticipated in the Introduction: while the cross sections are strongly enhanced for 
an $O(\rm MeV)$ temperatures, they are abruptly diminished to the primordial values as soon as the temperature drops below 
50 keV.

\FIGURE[h!]{
\includegraphics[scale=1]{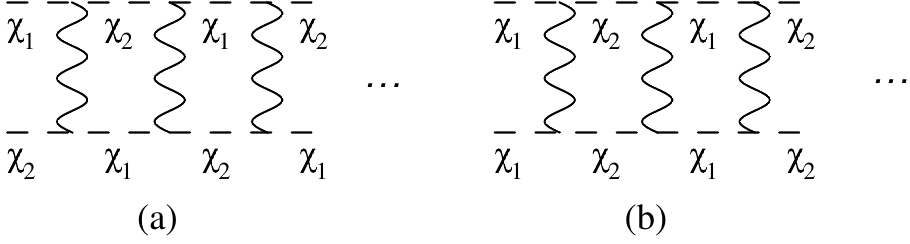}
\caption{Diagrams illustrates the formation of bound state.  
\label{fig:ladder}}}

\FIGURE[h!]{
\includegraphics[scale=1]{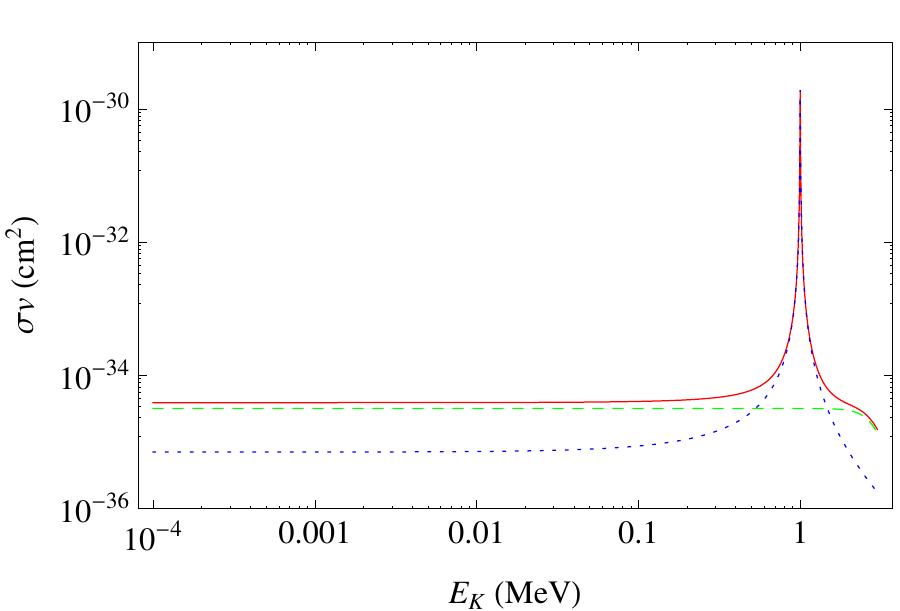}
\caption{$\sigma v$ as a function of kinetic energy of dark matter. The blue dotted and the green dashed curves show the contributions from the $\chi_1\chi_2$ resonance and the $\chi_1\chi_1$ annihilation 
\label{fig:sigma_v1}}}

\FIGURE[h!]{
\includegraphics[scale=1]{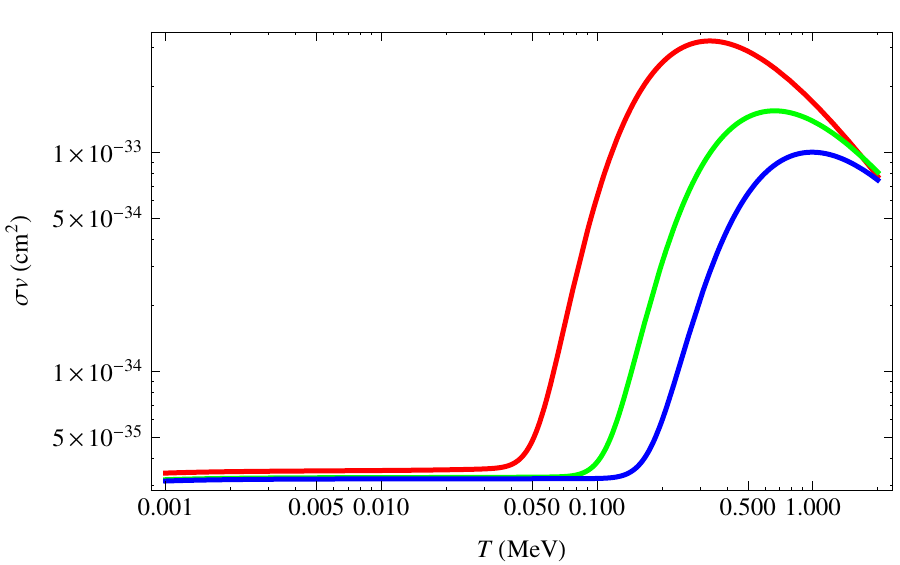}
\caption{$\langle\sigma v\rangle$ as a function of temperature. The red, green and blue curves are for $E_r = 0.5, 1 , 1.5$ MeV, respectively. 
One can observe a sharp drop-off in the averaged annihilation cross section below $T=50$ keV for all three cases.  
\label{fig:sigma_va1}}}


%

Finally we discuss the relation between $m_V$ and $\alpha_V$ that places $\chi_1\chi_2$ resonance just above 
the $\chi_2\chi_2$ separation threshold. It can be derived from the Schr${\ddot {\rm o}}$dinger equation 
\begin{equation}\label{schrodinger}
- \frac{\nabla^2}{2 \mu_\chi} \Psi + V(r) \Psi = E \Psi \ ,
\end{equation}
where $V(r) = - \frac{\alpha_V}{r} e^{-m_V r}$ and $\alpha = \frac{g^2}{4\pi}$, and $\mu_\chi \approx M_\chi/2$ is the reduced mass of the $\chi_1\chi_2$ system, and $E<0$ since it is a bound state
(for the discussion of $m_V-\alpha_V$ relation,  the transition to $\chi_2\chi_2$ continuum is a small perturbation and can be neglected). 

\FIGURE[h!]{
\includegraphics[scale=1]{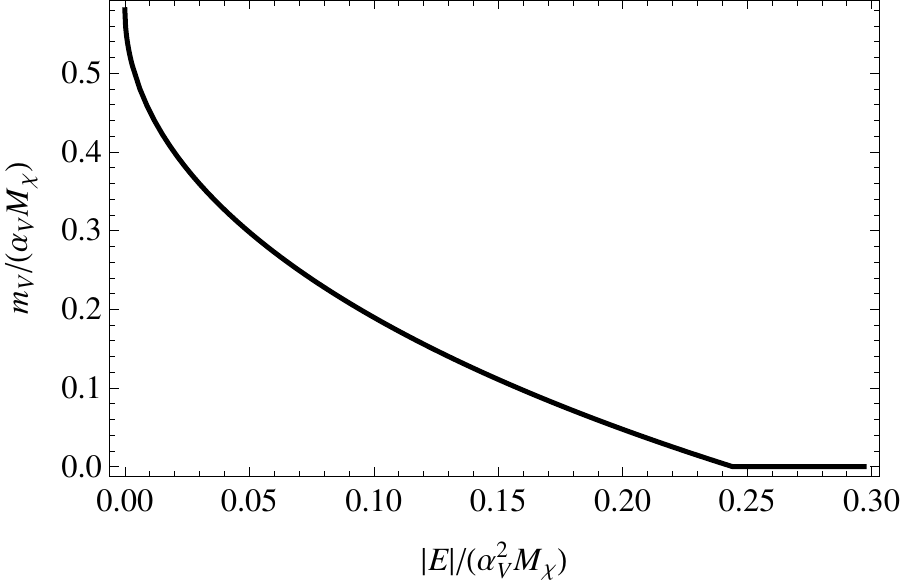}
\caption{Relation between $m_V/(\alpha_V M_\chi)$ and $|E|/(\alpha_V^2 M_\chi)$. 
\label{fig:get1s}}}

From the straightforward numerical analysis of this equation we derive 
the relation between $m_V/(\alpha_V M_\chi)$ and $E/(\alpha_V^2 m_V)$, shown in Fig.~\ref{fig:get1s}. 
The resulting domain 
of parameters is consistent with the requirement $m_V/(\alpha_V M_\chi) <0.6$ that we can obtain from our analysis of $\Gamma_{\rm in(out)}$.




\section{Narrow resonance in the $p$-wave}

\subsection{The model}

In this model we assume that the dark matter particle is a complex scalar interacting with two gauge bosons and the effective Lagrangian can be written as
\begin{equation}
{\cal L} = |\partial_\mu \chi - ig_1 V_{1\mu} \chi - ig_2 V_{2\mu} \chi|^2 \ .
\end{equation}
We would like $V_1$ to generate a 2$p$ resonance during the annihilation of dark matter when the kinetic energy is around 1 MeV, and we would like the mass of $V_2$ to be smaller than the energy difference between the 2$p$ resonance and the 1$s$ bound state so that the 2$p$ resonance can decay to 1$s$ state. However, in order for avoid overly large 
enhancement of the cross section due to light $V_2$ 
exchange constrained by the CMB, we need $g_2$ coupling to be small. 
Thus, the main idea behind this model is to realize the $p$-wave resonance-mediated 
capture into the  sub-threshold $1s$ state:
\begin{equation}
\chi+\chi^\dagger \to (\chi \chi^\dagger)_{2p~\rm resonance} \to (\chi \chi^\dagger)_{1s~\rm bound~state}+V_2\to {\rm light ~ mediators} \to {\rm SM}.
\end{equation}

In this model, in the early universe the main process for dark matter annihilation is 
\begin{equation}
\chi^\dagger \chi \rightarrow V_1 V_1 \ ,
\end{equation}
with the cross section given by
\begin{equation}
\label{relicab}
\sigma v = \frac{\pi \alpha_{1}^2}{M_\chi^2} \approx 2.3\times 10^{-26} {\rm cm^3/s} ~\Longrightarrow ~\frac{M_\chi}{\rm TeV} \simeq
\frac{\alpha_1}{0.025},
\end{equation}
where notation $\alpha_1 = 4g_1^2/(4\pi)$ is introduced. The relation between 
$M_\chi$ and $\alpha_1$ follows from the normalization to standard WIMP annihilation cross section.

\subsection{2$p$ bound state, $\Gamma_{\rm in}$ and  $\Gamma_{\rm out}$ }
At low energies the dynamics of a WIMP anti-WIMP pair is governed by the attractive Yukawa potential
$
V(r) = -\alpha_1 e^{- m_{V1} r}/r 
$. We analyze the  Schr${\ddot {\rm o}}$dinger equation in this potential numerically
imposing the condition of near-threshold $2p$ resonance. Our study shows that for $M_\chi=1$ TeV and  $\alpha_{V} = 0.025$ the choice of $m_V/(\alpha_V M_\chi) = 0.1201$ places this resonance at $E=1$ MeV. 
One can also quantify the degree of fine tuning required for having an MeV resonance by normalizing it on the binding energy of the 
$2p$ state in hydrogen-like case, $E_R/(\alpha_{1}^2 M_\chi/16) \sim (1-{\rm few})\%$. 

We then perform the study of the elastic scattering on this resonance to determine $\Gamma_{\rm in}$. 
The cross section for $p$-wave resonance scattering can be written as
\begin{equation}
\sigma_{\rm el} = \frac{12\pi}{k^2} \sin^2\delta = \frac{12\pi}{k^2} \frac{\Gamma_{\rm in}^2/4}{(E - E_R)^2 + \Gamma_{\rm in}^2/4}\ .
\end{equation}
with the explicit dependence of $\Gamma_{\rm in}$ on energy, 
$
\Gamma_{\rm in} = a E^{3/2} 
$.
The coefficient of proportionality $a$ can be calculated numerically, and for the same choice of 
parameters ($M_\chi=1$ TeV, $\alpha_1 = 0.025$,  $m_{V1}=3$ GeV) we determine $a = 1201$ GeV$^{-1/2}$. 
The resulting phase shift as a function of energy is  plotted in Fig. \ref{fig:sin2delta}.

\FIGURE[h!]{
\includegraphics[scale=1]{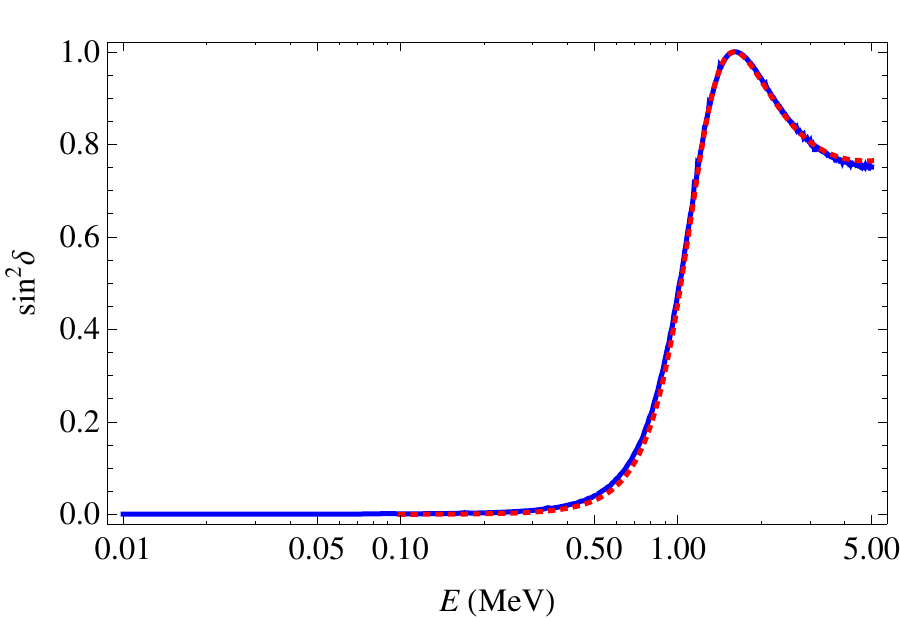}
\caption{The blue solid and red dashed curves show the numerical calculation and $\Gamma_{\rm in} = a E^{3/2} $ 
fit for $\sin^2\delta$ as a function of energy, 
for the choice of $M_\chi = 1000$ GeV, $\alpha_{1} = 0.025$, $m_{V1} = 3$ GeV.    
\label{fig:sin2delta}}}


We next address the out-going width that has two contributions, 
$\Gamma_{\rm out}= \Gamma_{ 2p~\rm annihilation} + \Gamma_{ 2p\to1s +V_2} $:
one from the annihilation of $\chi$ and $\chi^\dagger$ directly
in $p$-wave, and  the other from $2p\to 1s$ decay with emission of  $V_2$.
 The $1s$ WIMP-onium state of $\chi$ and $\chi^\dagger$ immediately decays, but its rate of decay is inconsequential 
for the size of the annihilation cross section. 

For the reference,  the the width of 1$s$ bound state is given by 
\begin{equation}
\Gamma_{\rm 1s} \approx (\sigma v)_0 4\pi |\psi(0)|^2 \approx 1.25 \alpha_{1}^3 {\rm ~GeV} \approx 20 ~{\rm keV} \ .
\end{equation}
Since the direct annihilation in a 2$p$ resonance is further suppressed compared to the above width  by a factor of 
$O(\alpha_{1}^2)$, we conclude that 
\begin{equation}
\Gamma_{ 2p~\rm annihilation} \leq10 ~{\rm eV}\ ,
\end{equation}
which is much smaller than what we need. 
According to the discussions in the previous section, in order to generate enough enhancement, $\Gamma_{\rm out}$ 
has to be around 1 keV. Therefore, decay width from the 2$p$ resonance to the 1$s$ bound state should be around 1 keV,
which is easy to achieve via 
\begin{equation}
\Gamma_{\rm out} \approx \Gamma_{ 2p\to1s +V_2} \approx \frac{1}{3}\left(\frac{2}{3}\right)^7 \alpha_{2} \alpha_{1}^4 M_\chi \ .
\end{equation}
After fixing the $\Gamma_{\rm out}$, we can use Eq.~(\ref{master}) to calculate $\sigma v$ around the resonance, which is shown in Fig.~\ref{fig:sigma_v_2}.

\FIGURE[h!]{
\includegraphics[scale=1]{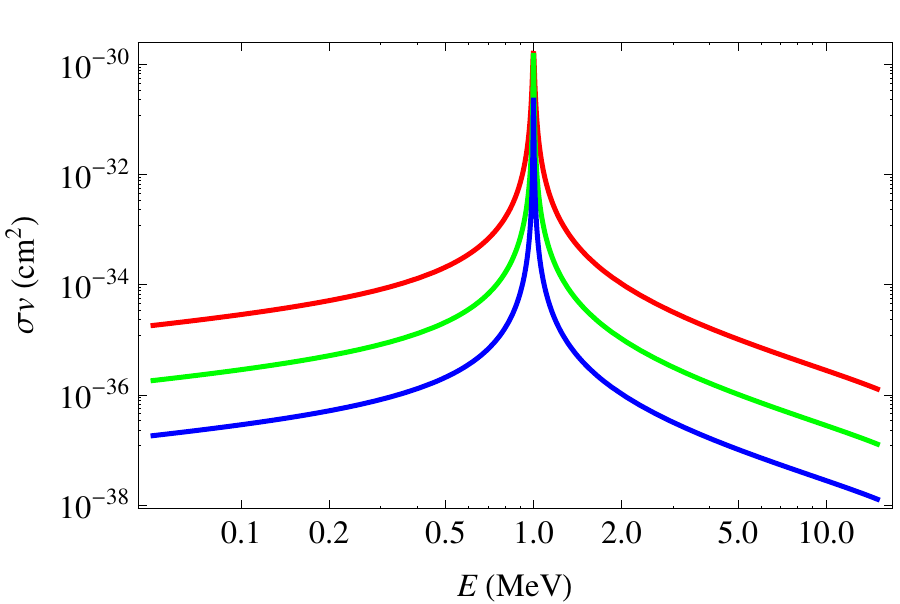}
\caption{$\sigma v$ around the $p$-wave resonance. The red, green and blue curves are for $\Gamma_{\rm out} = $ 10, 1, 0.1 keV, respectively.    
\label{fig:sigma_v_2}}}


From Eq.~(\ref{master}), we can further infer that in our case $\Gamma_{\rm out} \ll \Gamma_{\rm in}$, and
the total width is quite narrow at the resonance position, so that $\langle \sigma v\rangle$ can be written as
\begin{equation}\label{sigma_v_gamma_out}
\langle\sigma v\rangle_{\rm res} \sim \frac{\Gamma_{\rm out}}{M_\chi^3 v_0^4} \ ,
\end{equation}
where $v_0\approx10^{-3}$ is the average velocity of DM in the local galaxy.  Eq.~(\ref{sigma_v_gamma_out}) suggests that in order to get enough enhancement $\Gamma_{\rm out}$ should scale as $0.1\times(M_\chi/1 {\rm ~ TeV})^3$ keV. 
From the relic abundance relation (\ref{relicab}) we can further infer that
\begin{equation}\label{alpha2}
\alpha_{2} \sim 1.3\times10^{-2} \left( \frac{1~{\rm TeV}}{M_{\chi}} \right)^2 \ .
\end{equation}

Assuming again Maxwellian distribution, the averaged $\sigma v$ is shown in Fig.~\ref{fig:average}.
As the one can see, besides the $p$-wave induced resonance, the cross section also has a "flat part" associated with the Sommerfeld 
enhancement of the $s$-channel annihilation to be discussed below.

\FIGURE[h!]{
\includegraphics[scale=1]{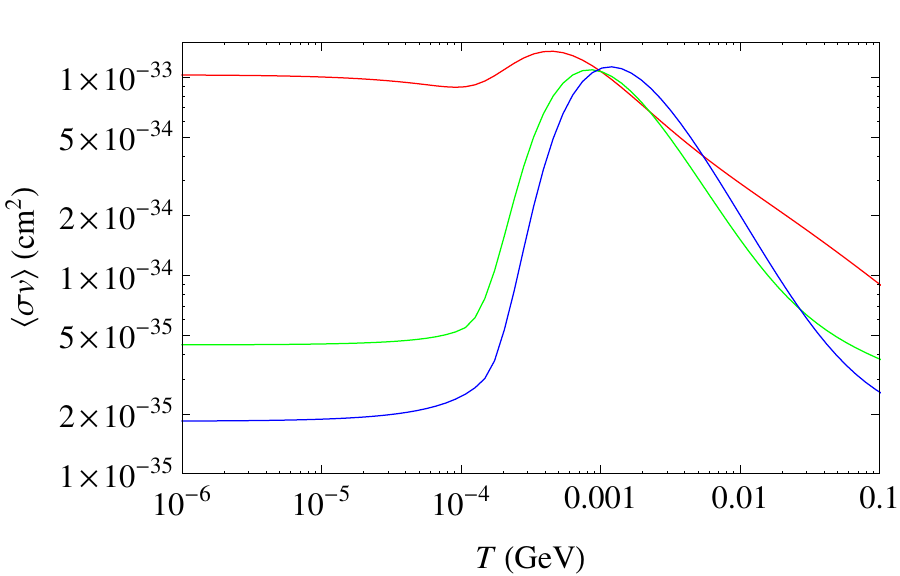}
\caption{Thermally averaged $\sigma v$ as a function of ``temperature'' assuming the velocity distribution is Maxwellian. The red, green and blue curves are for $M_\chi = $ 3000, 5000, 7000 GeV with $\Gamma_{\rm out} = $ 0.5, 3.3 and 10 keV, respectively. 
\label{fig:average}}}

\subsection{$s$-channel enhancement and constraints from CMB}

The vector boson $V_2$ may induce a large $s$-wave enhancement for the annihilation between $\chi$ and $\chi^\dagger$ when the kinetic energy is small, which will be strongly constrained by the observation of CMB. At the era of last scattering, the kinetic energy of the dark matter is much smaller than eV so that the dependence of the Sommerfeld enhancement factor on the kinetic energy of dark matter must be negligible~\cite{ArkaniHamed:2008qn}. Fig.~{\ref{fig:enhance_s}} shows the s-wave enhancement factor induced by $V_1$ and $V_2$ as a function of $M_\chi$ in the condition that $\alpha_{1}$ is fixed by the thermal relic abundance and $\alpha_{2}$ is determined by $\Gamma_{\rm out}$ from Eq.~(\ref{alpha2}). 

We quote recent interpretation of WMAP7 data as a limit on the dark matter annihilation cross section at the 95\% C.L. \cite{Galli:2011rz}:
\begin{equation}
f\frac{\langle\sigma v\rangle_{\rm CMB}}{M_\chi} < \frac{2.42\times10^{-27}{ ~\rm{cm^3/s}}}{\rm GeV} \ ,
\end{equation}
where $f$ is the energy deposition factor. Assumption of $f\sim1$ means that most of the energy from the annihilation of final states is efficiently transferred to the primordial plasma. Taking $f=1$, we show the CMB sensitivity 
by the blue line in Fig.~\ref{fig:enhance_s}. We can see that in the region with $M_\chi$ above 4 TeV 
the model is generally safe from the CMB constraints, {\em i.e.}
the  $2p$ resonance can enhance local annihilation,  
while the enhancement in the  $s$-channel is within limits. 


\FIGURE[h!]{
\includegraphics[scale=1]{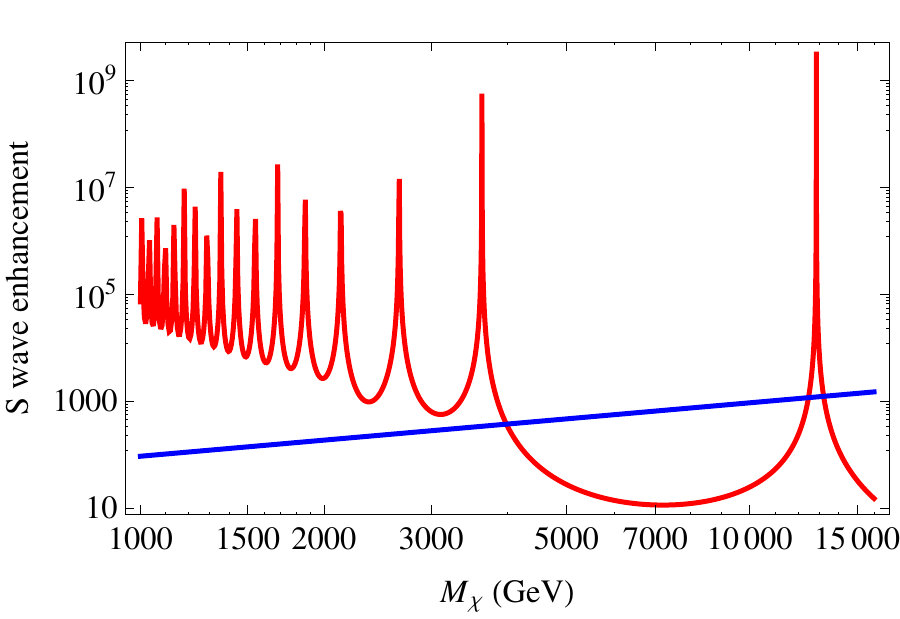}
\caption{The s-wave Sommerfeld enhancement factor induced by $V_1$ and $V_2$ as a function of $M_\chi$ with $\alpha_{V1}$ fixed by the requiring that the process $\chi\bar\chi\rightarrow V_1 V_1$ generates the right relic abundance of dark matter in the early universe by thermal annihilation and $\alpha_{V2}$ is determined from Eq.~(\ref{alpha2}).      
\label{fig:enhance_s}}}
%

\section{Constraint from annihilation in galactic center}

The models presented in the two previous sections avoid the CMB and dwarf spheroidal gamma constraints by design. 
This, however, does not exhaust all constraints on dark matter annihilation, and in this section we address constraints 
that usually come from (non)observation of gamma excess from the central region of our galaxy. For concreteness, 
we shall follow recent works that discuss implications of  H.E.S.S. data on high energy gamma-rays from the galactic centre (GC)~\cite{Abramowski:2011hc,Abazajian:2011ak}. One  familiar difficulty is that the density profile of dark matter in 
the GC is quite uncertain. In our model of dark resonance, this uncertainty is compounded 
by the velocity dependence of the cross section $\langle\sigma v\rangle$, which brings the sensitivity of the rates 
to the distribution of dark matter over velocities in GC. Given this velocity dependence of  $\langle\sigma v\rangle$, 
we argue that there is no conflict between these constraints and the enhancement factor for the 
DM annihilation within $\sim$1 kpc from the Solar System, relevant for the PAMELA signal explanation. 

The H.E.S.S. telescope is sensitive to high energy gamma-rays from a few hundred GeV to a few  tens of TeV, 
and can use its superior angular resolution to select  the gamma-ray signal from the GC, and compare its signal with the 
``background region''. 
The background region is chosen to be farther from the GC than the signal region in order to capitalize on the 
enhancement of the DM density  in the signal region relative to the background region,
inferred from numerical simulations of the DM profiles~\cite{Navarro:1996gj,Springel:2008by,Springel:2008cc}. 
Consequently, the average number of gamma-rays from the signal region 
should be larger than from the control region, 
and the difference between the two can be used for setting upper limits on the annihilation cross section. 

This technique is very sensitive to the assumptions about the density profiles, as emphasized in Ref.~\cite{Abramowski:2011hc}. 
For example,  a constant-density core of DM within the inner region of around 450 pc of the Milky Way will 
wash away 
all constraints that can be derived from HESS GC analysis. We now investigate how the velocity dependence of 
$\langle \sigma v\rangle$ will modify the conclusions of analysis in Refs.~\cite{Abramowski:2011hc,Abazajian:2011ak}. 
For concreteness, we shall assume a NFW density distribution for the Milky Way. 

The differential flux per solid angle generated by the DM annihilation can be written as
\begin{eqnarray}
\frac{d F}{d E} &=& \frac{1}{\Delta\Omega} \int_{\Delta\Omega}d\Omega \int_{\rm l.o.s} dx~ \rho^2(r_{\rm gal}(b,l,x)) \langle\sigma v\rangle \frac{1}{8\pi M_\chi^2} \frac{d N_\gamma}{d E} \nonumber\\
&=&\frac{\langle\sigma v\rangle_{\odot}}{2} \frac{J_{\Delta}^{\rm eff}}{J_0} \frac{1}{4\pi M_\chi^2} \frac{dN\gamma}{d E}  \ ,
\end{eqnarray} 
where $\rho$ is the energy density of DM, $r_{\rm gal} = \sqrt{R_\odot^2 - 2xR_{\odot} \cos(l)\cos(b) + x^2}$ is the distance between the line-of-sight point $x$ and GC, $J_0$ is the normalization constant defined as $J_0 \equiv \rho_{\odot}^2 d_{\odot}$, and 
\begin{equation}
J_{\Delta}^{\rm eff} = \frac{J_0}{\Delta\Omega} \int_{\Delta\Omega} d\Omega \int dx \rho^2(r_{\rm gal}(b,l,x)) \frac{\langle\sigma v (r_{\rm gal}(b,l,x))\rangle}{\langle\sigma v\rangle_\odot} \ .
\end{equation} 
The difference of our analysis from Refs.~\cite{Abramowski:2011hc,Abazajian:2011ak} is that here $\langle\sigma v\rangle$ also depends on $r_{\rm gal}$ through the dependence on the velocity distribution. 

The NFW profile is given by the expression
\begin{equation}
\rho(r) = \frac{\rho_0}{\frac{r}{R_s} (1+\frac{r}{R_s})^2} \ ,
\end{equation}
where for the Milky Way $R_s$ is about 20 kpc. Then the velocity dispersion can be calculated from 
Jeans equation and the numerical solutions is shown in Fig.~\ref{fig:nfw}. More details on the velocity 
structure of the central region can be found in \cite{Navarro:2008kc}. 
\FIGURE[h!]{
\includegraphics[scale=0.8]{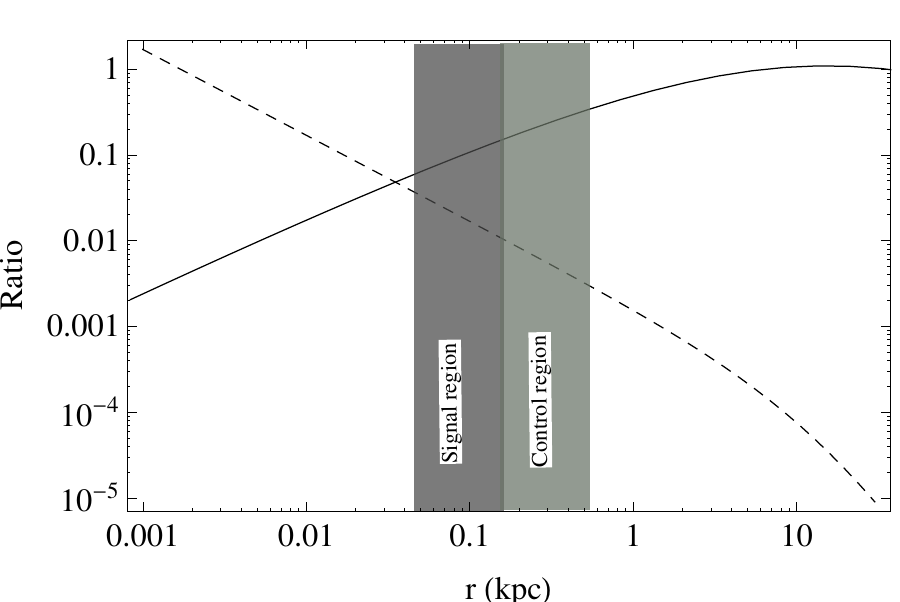}
\includegraphics[scale=0.8]{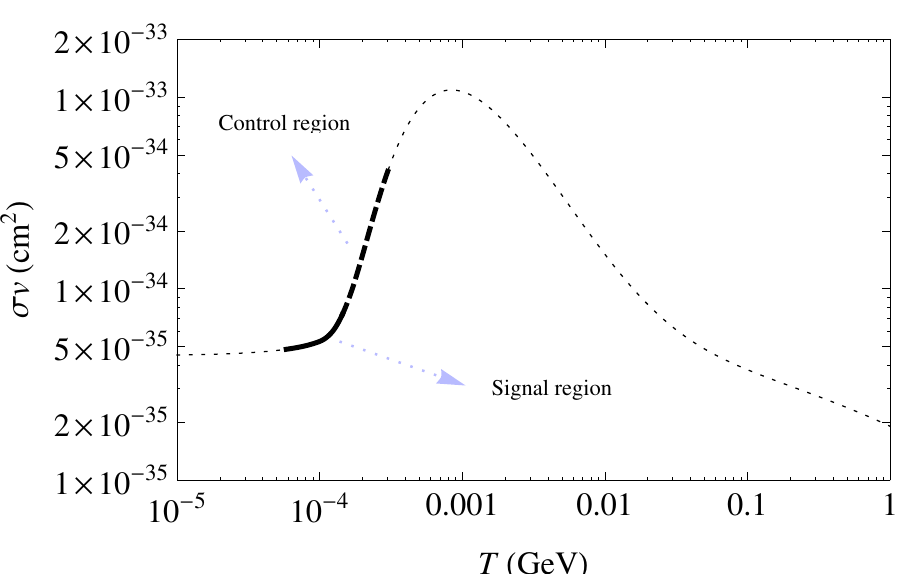}
\caption{Left: NFW distribution, the solid and dashed curves show the ratios of velocity dispersion and density distribution normalized to the same quantity near the Solar System. The two shaded regions are the signal region and control region, respectively. Right: the distribution of $\langle\sigma v\rangle$ in the signal region and control region in the second model with $M_\chi = 5000$ GeV, $\Gamma_{\rm out} = 3.3$ keV. The solid and dashed parts show the signal region and control region, respectively.  
\label{fig:nfw}}}
For simplicity, we further assume that the velocity distribution is Maxwellian.  Then the velocity dispersion, normalized on 
the dispersion around the Solar System, shown in Fig.~\ref{fig:nfw}, can be rigidly linked
 to the average of kinetic energy, and can be further interpreted in terms of effective temperature.
Therefore, from Fig.~\ref{fig:nfw}, we can see that in the signal region the effective temperature is reduced by about one order of magnitude whereas only by a factor of a few in the control region. 
Therefore, in models with the resonance "optimized" for the annihilation around the Solar System, 
the ratio $\bar J_\Delta^{\rm eff}({\rm signal})/\bar J_\Delta^{\rm eff}({\rm control})$ becomes 
smaller and the ensuing constraint is weaker. 
Since the information given in Ref.~\cite{Abramowski:2011hc} is not sufficient for an independent analysis, 
we only checked pixel 0 and pixel 2 in  Fig. 2 of Ref.~\cite{Abramowski:2011hc}, which are in the signal region and control region, respectively. Without considering the velocity dependence of $\langle\sigma v\rangle$ we get 
\begin{equation}
J_{\rm pixel~0} = 1015 J_0 \ , \;\;\; J_{\rm pixel~2} 
= 613 J_0 \ ,
\end{equation}
which are close to the average values in the signal region and control regions. Then, after including the velocity dependence
of the annihilation cross section, 
we can get
\begin{equation}
\bar J_{\rm pixel~0} = 161 J_0 \ , \;\;\; J_{\rm pixel~2} 
= 147 J_0 \ .
\end{equation}
Therefore, we can see that after considering the velocity dependence, the contrast between 
control and signal region is eroded, and 
the constraints from the H.E.S.S. GC gamma-ray observation become much weaker. 

\section{Summary and discussions}

Models with enhanced annihilation rate of dark matter aimed at explaining the 
PAMELA positron excess and Fermi-LAT $e^+e^-$ spectrum are strongly 
constrained by the absence of any evidence for non-thermal energy injection in the CMB  data, and by the 
absence of gamma-ray excess from dwarf spheroidal galaxies and the galactic centre by Fermi-LAT and H.E.S.S. respectively. 
In particular, models that have a Sommerfeld-type enhancement of annihilation at small velocities, that scales as inverse velocity
within some range, generally predict that the annihilation cross section $(\sigma v)_{\rm CMB,~dw.sph.} \geq
(\sigma v)_{\rm galaxy}$.
In this paper, we have shown that this is not the case in models 
where the annihilation proceeds via a  narrow resonance at kinetic energy around 1 MeV. 
Indeed, the narrowness of the resonance allows to maximize the cross section right in the desired range of 
WIMP kinetic energies. 

We propose two prototype models. In the first one, 
the mass splitting in the WIMP sector 
was used to create the $s$-resonance. The narrowness of the resonance, 
crucial for the success of this construction, 
comes from the fact that two sectors, $(\chi_2\chi_2)$ and $(\chi_1\chi_2)$ 
exist independently, and do not mix unless one introduces a new Higgs interaction that connect these two 
sectors. Consequently, the width of $\chi_1\chi_2$ resonance 
can tuned  to be around 1 keV,  which automatically shuts down the enhancement of $\sigma v$ 
when the kinetic energy of the DM particles becomes much smaller than an MeV. 
In the second model, we considered the $p$-wave resonance in the 
WIMP-aniti-WIMP state, so that the incoming width is proportion to $v^{3}$ where $v$ is
 the velocity of DM.  As a result, the low-velocity annihilation rate in this channel is automatically suppressed. 
It is fair to mention that these models are tuned, as the position of the resonance above the threshold is typically 
less than the energy level spacing in the bound state system, but we find that the degree of fine-tuning is $O(1\%)$, and 
far less if one were to achieve a similar resonance with the exchange by a heavy particle in the $s$-channel. 

The two models found here do not exhaust all possibilities, as the mechanisms for achieving the narrow resonance in the 
two-WIMP state are quite numerous. Another generic mechanism that we would like to mention here without building an 
explicit model for it goes as follows: the resonance can be achieved with the combination of attractive and repulsive force. 
For example, a Higgs and a vector exchange will create the following potential for the $\chi\chi$ system: 
$V_{\chi\chi} = (\alpha_V\exp(-m_Vr) -\alpha_h\exp(-m_hr) )/r$, while both pieces 
in the potential will be attractive for $\chi\chi^\dagger$. With $m_V<m_h$ and $\alpha_V < \alpha_h$, 
there will be a potential minimum at small $r$, separated from large $r$ by the barrier. Needless to say that this is exactly 
the case that can give narrow resonances, and the annihilation of $\chi\chi$ state via such resonance can be 
achieved via a doubly-charged Higgs state. 

Given that narrow resonances in the two-WIMP system can qualitatively 
change the outcomes of the DM annihilation constraints, 
we hope that the full analyses based on the formula (\ref{master}) with appropriate energy dependence of $\Gamma_{\rm in}$ 
will be performed in the near future. We close our paper with a series of additional comments:

\begin{itemize}

\item In our scenario, the annihilation cross section of DM peaked at MeV scale, which is about the 
starting temperature for the big bang nucleosynthesis (BBN). Thus one expect the enhancement of the 
annihilation rate during the BBN, with possible observable outcomes, depending on the specifics of the final state particles 
at the end of the annihilation cascade. We leave this subject for the future study~\cite{future}.

\item Diffuse galactic gamma ray constraint will provide a strong limit on the models discussed here, because it is less 
sensitive to the distribution over the velocity.
The most recent paper by FERMI collaboration \cite{Ackermann:2012rg} that appeared after our 
work has been completed, has shown to disfavour 
large enhancement of the annihilation rates for the TeV scale WIMPs. A few hundred  GeV WIMP could still be entertained as the 
PAMELA signal explanation, and we believe that with some modifications, our models with the near-threshold resonance 
can be made consistent with $O(100)$ GeV WIMP mass scale. 

\item While the enhancement of the lepton yield via a resonant channel was the primary 
reason for our investigation, we note that very similar resonant mechanisms can be invoked 
for enhancing the annihilation to monochromatic photons. The models of this type were previously 
discussed in \cite{Pospelov:2008qx}. 

\item Models discussed in this paper do have a considerable degree of self-interaction,
which can be constrained by  the ellipticity of DM halos~\cite{Feng:2009hw,Ibe:2009mk}, and the 
dynamics of the so-called "bullet" cluster~\cite{Markevitch:2003at}. 
The constraints from ellipticity is about two orders of magnitude stronger than from bullet cluster, and is given by
$
\frac{\sigma}{M_\chi} < 4.4\times10^{-27} ~{\rm cm^2~GeV^{-1}}
$.
For TeV dark matter, it is equivalent  to the sensitivity to  elastic 
cross section  at $\sim 10^{-24}$ cm$^2$, which is considerably larger than the unitarity bound in 
$l=0,1$ channels, thus not making any additional restrictions on models discussed here.

\end{itemize}

{\bf Acknowledgements.}  The authors would like to acknowledge useful conversations with Drs. J. Pradler, A. Ritz,  J. Zavala and P. Schuster. Research at the 
Perimeter Institute is supported in part by the Government of Canada through 
NSERC and by the Province of Ontario through MEDT.


\bibliographystyle{jhep}
\bibliography{resonance}

\end{document}